\documentclass[aps,prb,twocolumn,superscriptaddress,floatfix,longbibliography]{revtex4-2}

\usepackage{amsmath,amssymb}
\usepackage{bm}
\usepackage{graphicx}

\usepackage{textcomp}

\usepackage{enumitem}
\setlist{noitemsep,leftmargin=*,topsep=0pt,parsep=0pt}

\usepackage{hyperref}

\renewcommand{\vec}[1]{\bm{#1}}

\newcommand{\mytitle}{Computer Vision Applied to In-Situ Specimen Orientation Adjustment for Quantitative SEM Analysis}

\begin{document}

\title{\mytitle}

\author{Clay Klein}
\email[]{Clay.Klein@colorado.edu}
\affiliation{Department of Physics, University of Colorado, Boulder, Colorado 80309, USA}
\author{Chunfei Li}
\affiliation{Department of Physics, Clarion University of Pennsylvania, Clarion, Pennsylvania 16214, USA}

\date{\today}

\begin{abstract}
Quantitative analysis methods based on the usage of a scanning electron microscope (SEM), such as energy dispersive x-ray spectroscopy, often require specimens to have a flat surface oriented normal to the electron beam. In-situ procedures for putting microscopic flat surfaces into this orientation generally rely on stereoscopic methods that measure the change in surface vector projections when the surface is tilted by some known angle. Although these methods have been used in the past, there is no detailed statistical analysis of the uncertainties involved in such methods, which leaves an uncertainty in how precisely a specimen can be oriented. Here, we present a first principles derivation of a specimen orientation method and apply our method to a flat sample to demonstrate it. Unlike previous works, we develop a computer vision program using the Scale Invariant Feature Transform to automate and expedite the process of making measurements on our SEM images, thus enabling a detailed statistical analysis of the method with a large sample size. We find that our specimen orientation method is able to orient flat surfaces with high precision and can further provide insight into errors involved in the standard SEM rotation and tilt operations.
\end{abstract}

\maketitle
\section*{\label{sec:Introduction}Introduction}

Quantitative analysis methods frequently used in scanning electron microscopy (SEM) generally must account for the effects of geometric features on the specimen being analyzed. In particular, specimen porosity and surface topography in the region of interest can significantly affect the results of quantitative analysis methods \cite{Ronnhult1987, Winiarski2021}, and properly accounting for these geometric effects is an essential part of any analysis when reliable results are desired; accounting for surface topography often requires the adjustment of the specimen’s orientation relative to the electron beam, so this has become an essential part of specimen preparation. When the specimen is prepared outside the SEM or can be seen with the naked eye, the problem of specimen orientation is easily solved because the specimen can be directly placed into the necessary orientation \cite{Pohl2010, Webber2020}. However, if a specimen must be analyzed as is or the surface of interest cannot be seen outside the SEM, an in-situ method to orient the specimen must be employed. For a typical modern SEM, the only information and degrees of freedom available during an in-situ specimen orientation are SEM images and the rotation and tilt operations of the SEM stage, respectively.

Energy dispersive x-ray spectroscopy (EDS) is one example of an analysis method where specimen geometry is especially important. With EDS, obtaining quantitative data requires that a specimen be analyzed at a sufficiently flat surface oriented normal to the electron beam and that specimen porosity and topography in the surrounding region do not influence the emission or detection of x-rays \cite{Goldstein2003, Newbury2013a, Newbury2013b}. These constraints originate from the fact that the data obtained from specimens of interest are compared with data taken from a standard flat and non-porous specimen \cite{Jones2021}. While small errors in the orientation of a specimen relative to the electron beam may only have a small effect on the detection of x-rays \cite{Lifshin2001}, it is important to be able to place bounds on the errors and have a quantitative understanding of how large these errors are. Electron backscatter diffraction (EBSD) also has a geometric constraint that must be met for optimal data collection. In this case, a flat surface must generally be tilted $70^{\circ}$ towards the EBSD detector \cite{Schwartz2009}, and previous applications of EBSD have had to account for this requirement \cite{Slavik1993}. These two examples, along with other applications where specimen orientation is important, motivate the development of simple, accurate, and precise in-situ methods for orienting specimens in the SEM.

Stereoscopic methods have been used in the past to orient flat surfaces on specimens in a SEM so that a quantitative analysis can be performed. The first application of such a method to EDS appears to be by Bomback in 1973 \cite{Bomback1973}. Additionally, stereoscopic methods have been used in microscopy for other purposes, such as constructing three-dimensional models of SEM specimens \cite{Hena2018}, analyzing cleavage plane orientations \cite{Chen1991}, and consistently orienting microfossils \cite{Macleod1984}, as well as other applications \cite{Hilliard1972, Stevens1983, Themelis1990}. Although these methods have been used, there is no statistical analysis of the precision with which a specimen can be oriented, so there consequently remains an uncertainty in quantitative data when such a method has been used. Moreover, methods in the past have only shown use of one measurement for the determination of the angles required to orient a specimen; this small number of measurements does not provide sufficient data for a statistical analysis. Errors such as these have not been investigated in the past, so there remains a shortcoming in regards to how precisely a specimen can be oriented with these methods.  A quantitative analysis of these errors is therefore necessary to develop confidence in stereoscopic methods for orienting specimens.

In this paper, we develop an efficient and precise method for orienting flat specimens in the SEM such that the surface normal is aligned with the electron beam. We first provide a derivation of the method from first principles and then continue by applying the method to a dataset obtained from a macroscopic flat SEM specimen. To facilitate a detailed statistical analysis of the method and take advantage of the progress of computer technology, we develop and apply a computer vision program to obtain a large number of measurements on our SEM images. We find that the method is able to orient specimens with high precision and that the use of the program is not only very precise, but also more convenient for the SEM operator since it eliminates the need for manual measurements. Furthermore, we describe how our method was able to reveal errors present in the SEM stage or operations and provide a suggestion on how to minimize these errors.

\section*{\label{sec:Materials and Methods}Materials and Methods}

\subsection*{Specimen Orientation Method}

\begin{figure*}[ht]
    \includegraphics[clip=true,width=2\columnwidth]{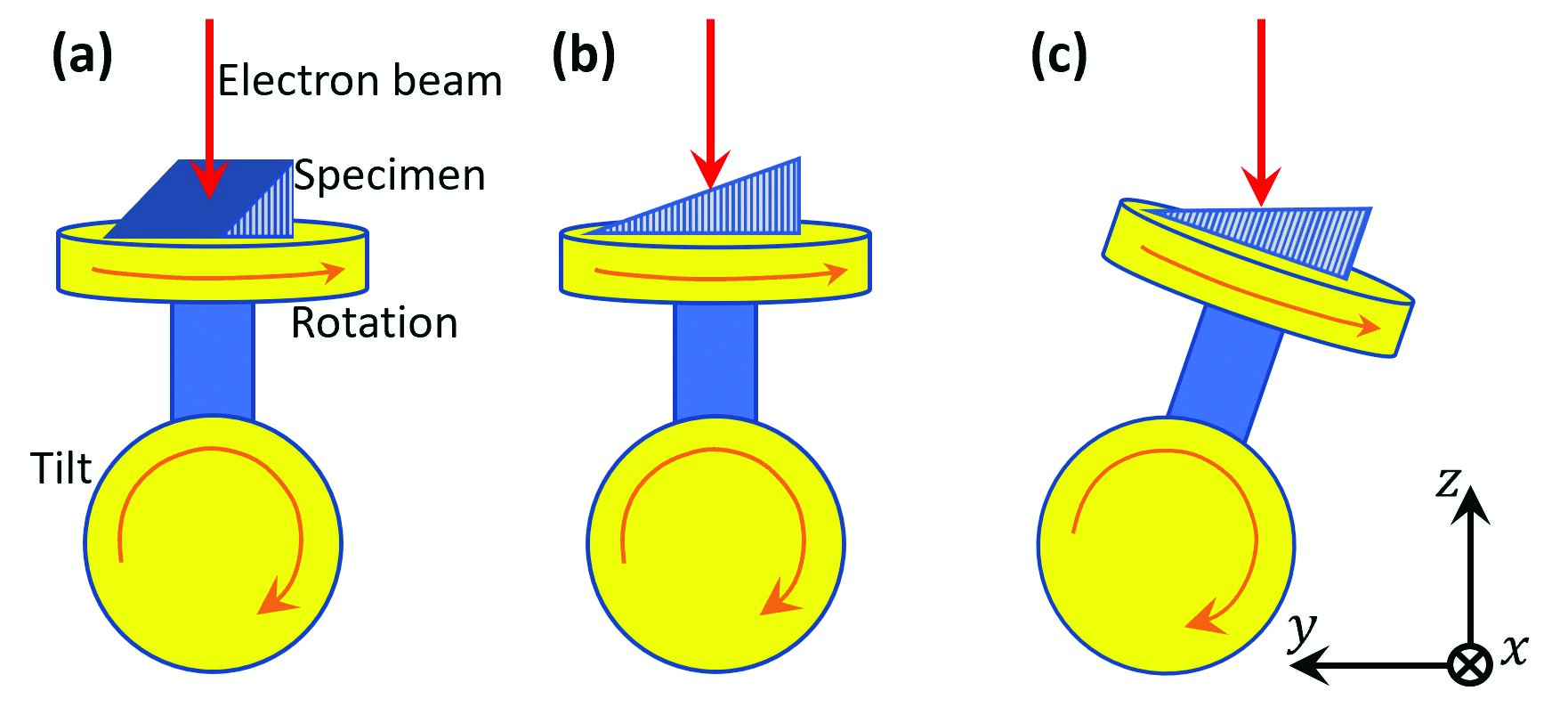}
    \caption{A schematic drawing of a typical SEM stage and the coordinate system. A flat plane on a specimen
must be rotated from its initial orientation (a) to an intermediate stage (b), where it can then be tilted so
that the plane is oriented with its surface normal aligned with the electron beam (c).}
     \label{fig:fig1}
\end{figure*}

The SEM used in the present case has two operations that enable in-situ adjustment of a specimen’s orientation: rotation and tilt. The rotation operation rotates the specimen through an angle $\phi$ about an axis parallel to the electron beam; we define this axis as the $z$-axis. The tilt operation tilts the specimen through an angle $\theta$ about an axis that we define as the $x$-axis. The $y$-axis is orthogonal to both the $x$- and $z$- axes and follows the right-hand rule. In order to orient a flat plane on a specimen, we must first rotate the SEM stage through an angle $\phi_0$ and then tilt it by an angle $\theta_0$, as shown in Fig. \ref{fig:fig1}. We define the proper angles as the pair of rotation and tilt angles $(\phi_0, \theta_0)$ that orient the plane such that its surface normal is parallel to the $z$-axis, as in Fig. \ref{fig:fig1}(c).

The proper angles may be determined by solving a system of equations involving measurements made on two SEM images taken at different stage tilt angles. We describe the mathematical details of the system of equations in the following.

Using the aforementioned coordinate system, the rotation and tilt operations in the SEM may be described by rotation and tilt operators in the canonical basis, which are respectively given by

\begin{equation*}
    \textbf{R}_\phi = 
    \begin{bmatrix}
        \cos \phi & -\sin \phi & 0\\
        \sin \phi & \cos \phi & 0\\
        0 & 0 & 1
    \end{bmatrix}
\end{equation*}
and
\begin{equation*}
    \textbf{T}_\theta = 
    \begin{bmatrix}
        1 & 0 & 0\\
        0 & \cos \theta & -\sin \theta\\
        0 & \sin \theta & \cos \theta
    \end{bmatrix}.
\end{equation*}
These operators rotate and tilt vectors by $\phi$ and $\theta$ in the same way that the SEM stage does.

Consider a plane on the specimen which is properly oriented; this plane can be modeled by the unit normal vector $\hat{\bm{k}} = \begin{bmatrix} 0 & 0 & 1 \end{bmatrix}^{\textrm{T}}$, where $\textrm{T}$ is the transpose operation. Apply the operators as follows to tilt and rotate $\hat{\bm{k}}$ by $-\phi_0$ and $-\theta_0$ to obtain the unit normal vector $\hat{\bm{u}}$ of a plane with proper angles $(\phi_0, \theta_0)$:

\begin{equation*}
    \hat{\bm{u}} = \textbf{R}_{-\phi_0}\textbf{T}_{-\theta_0}\hat{\bm{k}} = 
    \begin{bmatrix}
        \sin \phi_0 \sin \theta_0 \\
        \cos \phi_0 \sin \theta_0 \\
        \cos \theta_0
    \end{bmatrix}.
\end{equation*}

Suppose we have a vector $\vec{v}_i$ on the plane with components given by $\vec{v}_i = \begin{bmatrix} x_i & y_i & z_i \end{bmatrix}^{\textrm{T}}$. Since $\vec{v}_i$ is a vector on the plane and $\hat{\bm{u}}$ the unit normal vector of the plane, we may use the condition $\vec{v}_i \cdot \hat{\bm{u}} = 0$ to parametrize $z_i$ in terms of $x_i$, $y_i$, $\phi_0$, and $\theta_0$ as follows:

\begin{align*}
    0 &= \vec{v}_i \cdot \hat{\bm{u}}\\
    &= x_i \sin \phi_0 \sin \theta_0 + y_i \cos \phi_0 \sin \theta_0 + z_i \cos \theta_0\\
    \Longrightarrow z_i &= -x_i \sin \phi_0 \tan \theta_0 -y_i \cos \phi_0 \tan \theta_0.
\end{align*}
Thus, the set of all vectors on the plane is given by

\begin{equation*}
    \left\{ 
    \begin{bmatrix}
        x_i \\
        y_i \\
        -x_i \sin \phi_0 \tan \theta_0 -y_i \cos \phi_0 \tan \theta_0
    \end{bmatrix} 
    \middle| x_i, y_i \in \mathbb{R}
    \right\};
\end{equation*}
this expression enables us to write any vector on the plane in terms of the $x$ and $y$ components and the proper angles.

The key to orienting a plane under the constraints of the SEM lies in observing how the vectors change when tilted by an angle $\Delta \theta$ chosen by the SEM operator:

\begin{widetext}
\begin{equation*}
    \textbf{T}_{\Delta \theta} \vec{v}_i = 
    \begin{bmatrix}
        x_i \\
        y_i \cos \Delta \theta + \sin \Delta \theta \tan \theta_0 \left( x_i \sin \phi_0 + y_i \cos \phi_0 \right) \\
        y_i \sin \Delta \theta - \cos \Delta \theta \tan \theta_0 \left( x_i \sin \phi_0 + y_i \cos \phi_0 \right)
    \end{bmatrix}  = 
    \begin{bmatrix}
        x_i \\
        y_{iT} \\
        z_{iT}
    \end{bmatrix}.
\end{equation*}
\end{widetext}

This matrix equation represents three separate equations. $x_i$, $y_i$, and $y_{iT}$ are quantities that can be measured from the SEM images, and $\phi_0$, $\theta_0$, $z_{iT}$ are unknown. Our purpose is to solve for $\phi_0$, and $\theta_0$. The first equation is trivial, and the third one involves the unknown $z_{iT}$, which we are not interested in. The second equation involves the two unknown quantities we desire in terms of known or measurable quantities. Thus, if we consider two vectors represented by the subscript $1$ and $2$, we obtain the following system of equations:
\begin{align}
\label{eqn:eqn1}
\begin{split}
    y_{1T} &= y_1 \cos \Delta \theta + \sin \Delta \theta \tan \theta_0 \left( x_1 \sin \phi_0 + y_1 \cos \phi_0 \right) \\
    y_{2T} &= y_2 \cos \Delta \theta + \sin \Delta \theta \tan \theta_0 \left( x_2 \sin \phi_0 + y_2 \cos \phi_0 \right).
\end{split}
\end{align}
This system of equations can be solved for the proper angles $(\phi_0, \theta_0)$ if the two vectors chosen are linearly independent.

\subsection*{Computer Vision}

Measurements for $x_i$, $y_i$, and $y_{iT}$ were obtained by making use of the Scale Invariant Feature Transform (SIFT) \cite{Lowe2004}. A custom-built Python program using the SIFT method in the OpenCV library was created which identifies a number $n$ of matched points in two SEM images. The system of equations was solved using the least squares method in the SciPy library. From the $n$ matched points,  $m=C_2^n$ unique vectors can be formed, from which $C_2^m$ unique pairs of vectors can be obtained. Ideally, the system of equations (\ref{eqn:eqn1}) could be solved for all pairs of vectors, however in practice we must apply a set of constraints to obtain accurate results. The constraints applied in the present case are as follows: (a) the change in the $x$ coordinate between images for each point is close to $0$, (b) the change in $y$ coordinate between images for each point is bounded, (c) the projected vectors are sufficiently long, and (d) the angle between the projected vectors for each vector pair is sufficiently large. (a) and (b) serve to eliminate points which are incorrectly matched, and (c) and (d) are essential to ensure that the length and angle quantities are not comparable to deviations which may occur due to surface roughness and other small factors of variability.

The exact numerical values that are used in these constraints will depend on the parameters of the SEM being used. In our case, we had SEM images of size $828$ by $768$ pixels. The change in $x$ coordinate between matched points for the two images was restricted to $1$ pixel, and the change in $y$ coordinate was restricted to $100$ pixels. Matched points which do not meet these constraints were discarded. Additionally, the length of the vectors was required to be at least $300$ pixels, and for each pair of vectors the angle between the vectors was required to be between $30^{\circ}$ and $150^{\circ}$. Pairs of vectors that do not meet these constraints were discarded.

\subsection*{SEM Dataset}

A TESCAN Vega-3 SEM was used for imaging. The SEM stage is capable of rotating $360^{\circ}$ and tilting approximately $\pm 30^{\circ}$ depending on the stage position. A transmission electron microscope grid covered with Formvar is used as a specimen, and the specimen is glued to a pre-tilted SEM specimen holder with the Formvar side up. The area analyzed is part of the frame to ensure flatness and no deformations. The Formvar provided sufficient features for the computer vision program to recognize. SEM images were taken at rotation angles from $0^{\circ}$ to $350^{\circ}$ in increments of $10^{\circ}$ and at tilt angles of $0^{\circ}$, $20^{\circ}$, and $-20^{\circ}$. Each image was taken at a $50$ $\mu$m field of view with the same point being at the center of the image in each case. The specimen was oriented in such a way that the proper rotation and tilt angles were both nonzero.

\begin{figure*}[ht]
    \includegraphics[clip=true,width=2\columnwidth]{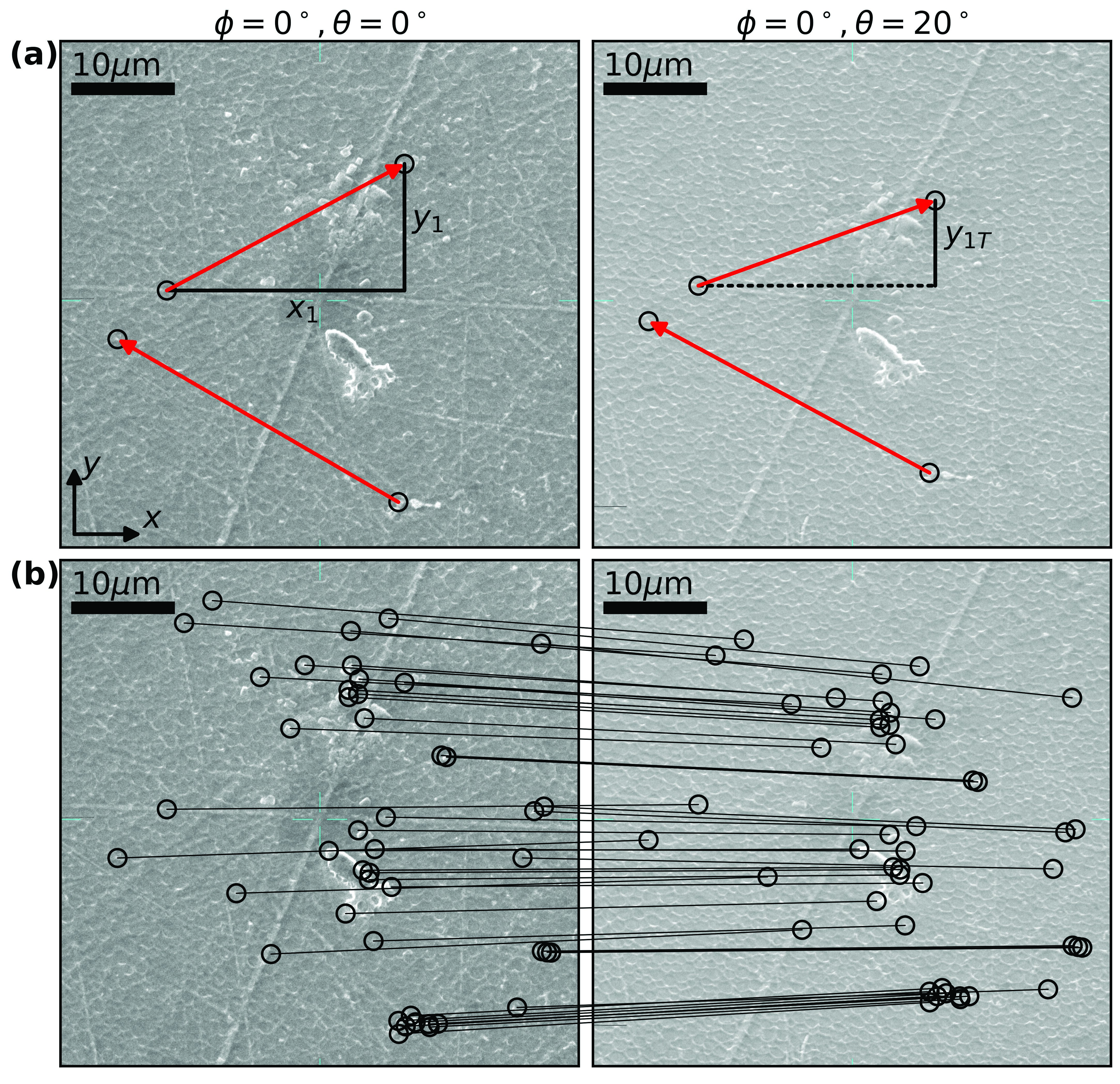}
    \caption{(a) One pair of vectors that was identified from a sample of four matched points and (b) all of the
49 matched points identified for this particular image pair. The drawings on (a) show how the $x$ and $y$
components of a vector are determined, although they are automatically calculated by the program in
reality.}
     \label{fig:fig2}
\end{figure*}

\section*{Results}

\begin{figure*}[ht]
    \includegraphics[clip=true,width=2\columnwidth]{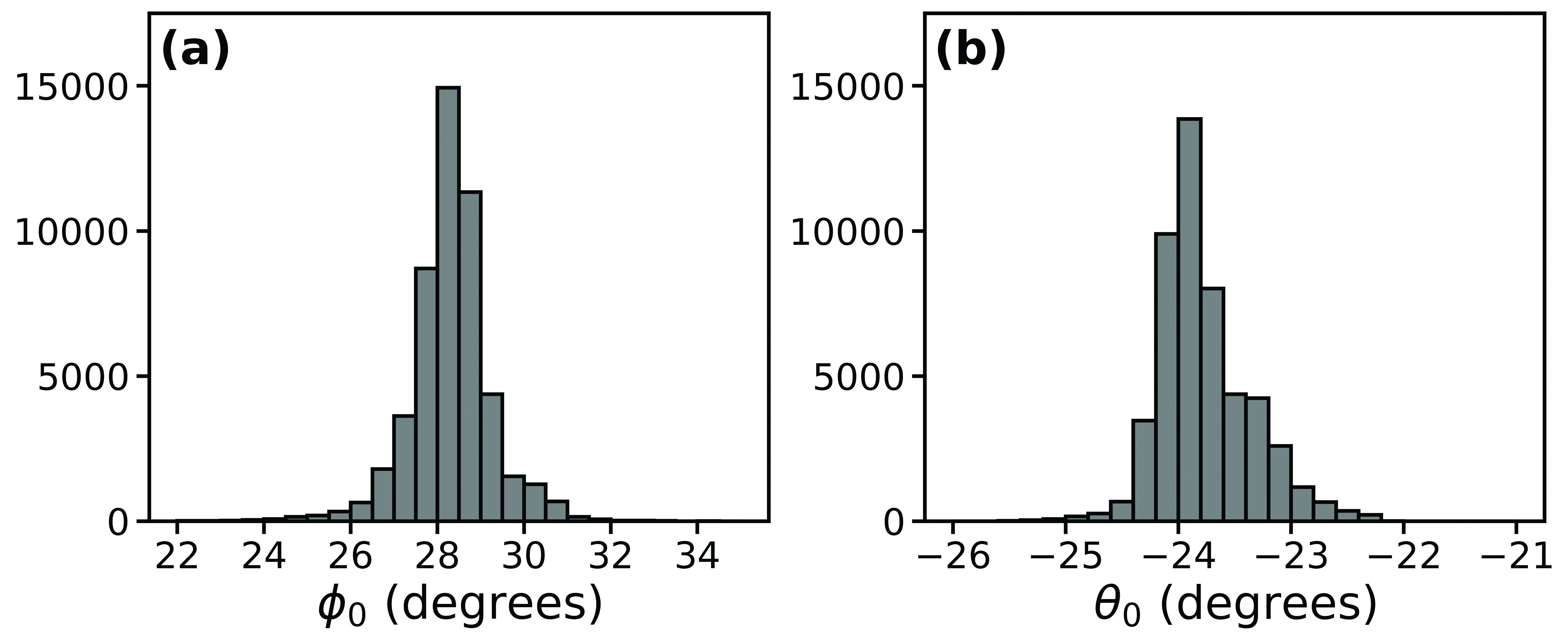}
    \caption{The distribution of the proper (a) rotation and (b) tilt angles calculated using pairs of vectors
formed from the matched points shown in Fig. \ref{fig:fig2}(b).}
     \label{fig:fig3}
\end{figure*}

As a first example and test of our program, we applied it to a pair of two images separated by a tilt angle of $\Delta \theta = 20^{\circ}$. $49$ matched points were identified, from which $690,900$ possible pairs of vectors can be formed. However, after applying the aforementioned constraints, this was reduced to $50,120$ pairs of vectors, approximately $7$\% of all possible pairs. A sample of $4$ points identified by the program and one possible pair of vectors as well as all matched points are shown in Fig. \ref{fig:fig2}. The system of equations (\ref{eqn:eqn1}) was solved for all $50,120$ pairs of vectors meeting the constraints to obtain $50,120$ proper angles. The distributions of the calculated proper angles are shown in Fig. \ref{fig:fig3}. A relatively narrow and approximately Gaussian distribution of proper angles with a clear peak is apparent, although there is a slight asymmetry in the proper tilt angles in Fig. \ref{fig:fig3}(b). The median and standard deviation of the proper rotation angles $\phi_0$ are $28.3^{\circ}$ and $1.0^{\circ}$, while the median and standard deviation of $\theta_0$ are $-23.9^{\circ}$ and $0.4^{\circ}$, respectively.

\begin{figure*}[ht]
    \includegraphics[clip=true,width=2\columnwidth]{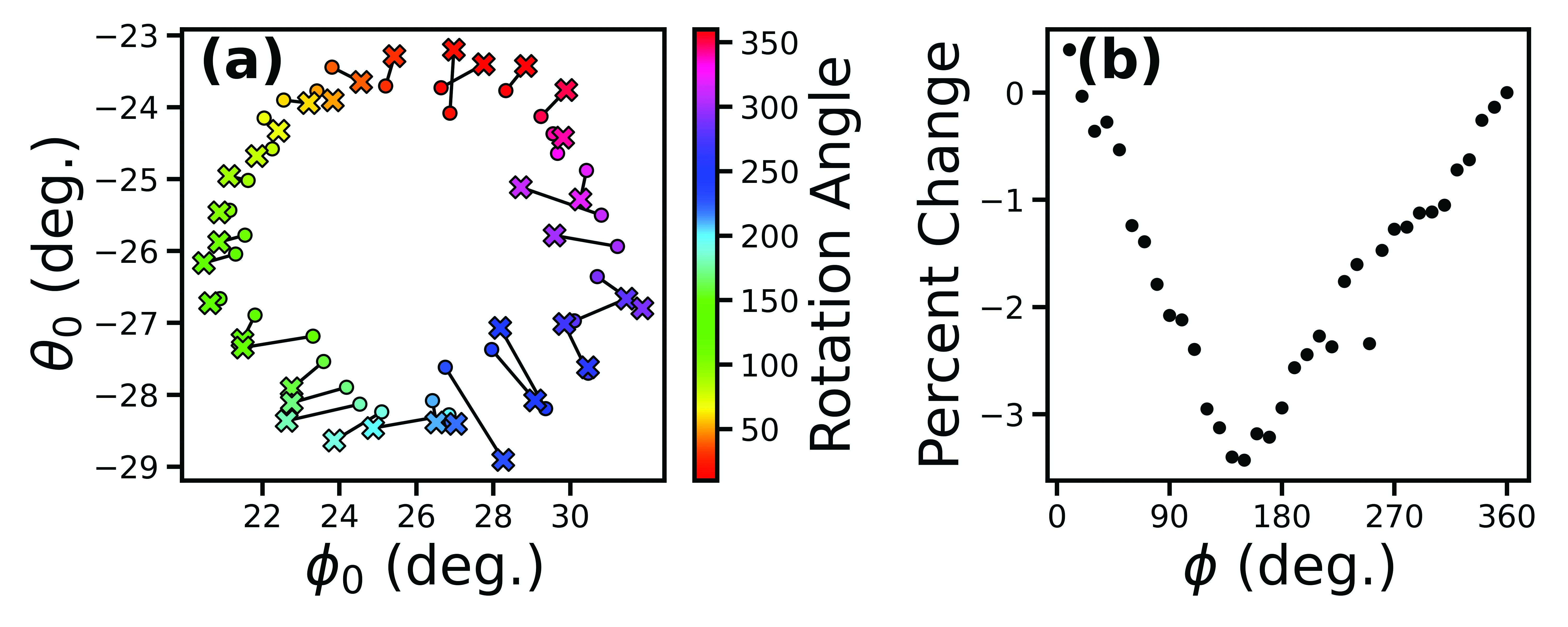}
    \caption{(a) The proper angles calculated at each rotation angle; dots denote those calculated using a tilt angle of $\Delta \theta = 20^{\circ}$ while an X denotes $\Delta \theta = -20^{\circ}$, and a line connects points at the same rotation angle. (b) The percent change of the distance between two points relative to the distance at $\phi=0^{\circ}$. The SEM’s error function shown in (b) leads a range of proper angles in (a).}
     \label{fig:fig4}
\end{figure*}

In addition to the two images considered above, we also used the program to evaluate the proper angles at each rotation angle from $0^{\circ}$ to $350^{\circ}$ in increments of $10^{\circ}$. For each rotation angle, we used a tilt angle of $\Delta \theta = 20^{\circ}$ and $\Delta \theta = -20^{\circ}$ to pair with the $0^{\circ}$ tilt images. The benefit of this analysis is that it provides a diverse set of images to test the program on and further enables us to characterize the variance of the resulting proper angles to understand how precise the method is. We expect the proper angles calculated from all of the image pairs to follow a Gaussian distribution in the $\phi$-$\theta$ parameter space because no other factors of variability in the proper angles are expected other than statistical fluctuations in the measurements. However, as shown in Fig. \ref{fig:fig4}(a), the proper angles follow a circular pattern which correlates with the rotation angle of the SEM. As described in the Discussion, this can be explained by a slight error in the SEM stage or specimen stub; if the SEM stage or specimen stub tilts slightly as the SEM stage is rotated, then it can induce changes in the proper angles. This idea is confirmed by the data in Fig. \ref{fig:fig4}(b), which shows the percent change in the apparent distance between two points as a function of rotation angle while the tilt angle is set to $0^{\circ}$. Because the stage rotation axis is supposed to be parallel to the electron beam, the distance between two points on the SEM images are expected to be constant as the stage is rotated. However, since this distance changes systematically as the specimen is rotated, we conclude that the specimen is tilting slightly in a way that depends on the rotation angle.

\section*{Discussion}

The research presented here expands upon our previous work \cite{Klein2022}, in which we demonstrated a different method for orienting specimens in the SEM. Both projects are based on the idea that changes between SEM images taken at different tilt angles depends on the proper angles of the specimen, and an analysis of these changes can be used to determine the proper angles. A brief description of the previous work is provided here, and its weaknesses, which have been improved upon in the present work, are analyzed. In the previous work, we focused on the projected distance between two points of the specimen surface, the line between which is perpendicular to the SEM tilting axis; this distance can be measured directly on the SEM images. At a fixed stage rotation angle, this distance was measured twice on SEM images corresponding to two different tilt angles, and the ratio of these measurements was calculated. The same procedure was repeated for different rotation angles and the resulting ratios plotted as a function of the stage rotation angle. Fitting the experimental data with a mathematical model then enables us to extract the proper rotation and tilt angles. While our previous work was successful as a proof of concept, it is not practical for the following reasons. First, a large amount of SEM images must be taken, which hinders its useability in places such as service labs. At a step size of $10^{\circ}$, 72 images must be taken to obtain a complete curve. Higher accuracy requires a smaller step size and thus more SEM images. Second, the measurements in our previous work were carried out manually, which leaves room for uncertainty due to subjective judgement. Such measurements and the data processing that follows are also very labor intensive. In contrast, the present work requires only two images, and all measurements and calculations are done automatically. A user of the program needs only to provide two SEM images as an input, and the proper angles are generated as the output.

Our analysis shows that when using two images to orient a specimen, a range of proper angles can be obtained which have an approximately Gaussian distribution (Fig. \ref{fig:fig3}). Although the range is relatively narrow, the accuracy of the specimen orientation method can be improved by taking the mean or median of the proper angles, thus eliminating statistical fluctuations which may occur in individual measurements. The measurements used to calculate the proper angles are all valid measurements that one might reasonably make by hand when orienting a specimen, so by considering a large number of measurements, we therefore obtain an understanding of all the possible values that one might obtain for the proper angles.

\begin{figure}[ht]
    \includegraphics[clip=true,width=\columnwidth]{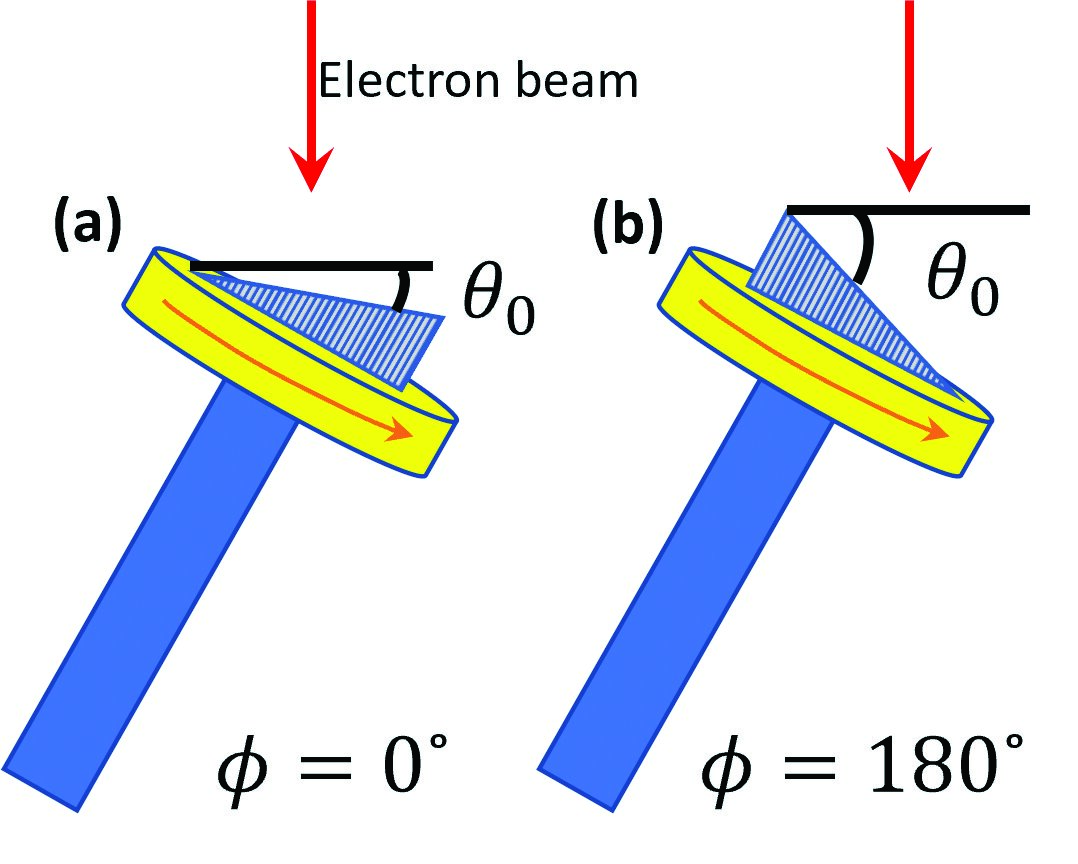}
    \caption{A specimen rotated around an axis that is not parallel to the electron beam. The proper tilt angle
at a rotation angle of $0^{\circ}$(a) is different than at a rotation angle of $180^{\circ}$(b).}
     \label{fig:fig5}
\end{figure}

The variance in the proper angles for one pair of images is quite small. However, we also used our program to calculate the proper angles for each rotation angle from $0^{\circ}$ to $350^{\circ}$ in increments of $10^{\circ}$ using two different tilt angles for each rotation angle. In this case, we obtained a circular pattern for the proper angles in the $\phi$-$\theta$ parameter space (Fig. \ref{fig:fig4}). We believe that each of the proper angles shown in Fig. \ref{fig:fig4} are correct, but that the proper angles of the specimen change as the SEM stage is rotated. This could result if the SEM rotation axis is not perfectly parallel to the electron beam; as shown by the simplified and exaggerated schematic drawing in Fig. \ref{fig:fig5}, a rotation of the specimen would change the proper angles if this were the case. This form of error would change both the proper tilt and rotation angles if the true rotation axis is on neither the $x$-$z$ nor $y$-$z$ plane, as is believed to be the case for our SEM. As pointed out by Lifshin and Gauvin \cite{Lifshin2001}, the fact that an SEM stage is not perfectly constructed can induce errors in a quantitative analysis because the true rotation and tilt angles of the SEM stage may not match the stated values. The data in Fig. \ref{fig:fig4} presents a quantitative demonstration of this form of error and its effect on the calculation of the proper angles.

When a quantitative analysis of a specimen in a SEM is desired and use of an in-situ specimen orientation method is necessary, it is important to take into account errors such as these. For the type of error we encountered here, we suggest a simple procedure which will optimally orient a flat specimen such that its surface normal is aligned with the electron beam. After performing a specimen orientation method and orienting the specimen at the calculated proper angles, the procedure can be performed again, with the previous proper angles as the new reference point; if no error is present, then the calculations will indicate that the proper angles relative to this new reference point are approximately $(0^{\circ}, 0^{\circ})$. If they are non-zero, then orienting the specimen again to the newly calculated proper angles will achieve nearly optimal accuracy. This can significantly increase the accuracy of the method while only requiring the inconvenience of doing the procedure twice instead of once. Additionally, it is important that the surface under consideration be flat in the entire field of view of the SEM image. Small undulations on the surface of an apparently planar surface can induce errors in the calculation of the proper angles. If a large number of points are identified on such a surface, then we would expect a larger standard deviation in the calculation of the proper angles; the average value of the proper angles in this case may orient a plane which is fitted to the average specimen surface rather than any specific portion of the specimen.

The computational efficiency and robustness of the program we used to perform our analyses are also relevant for the practicality of the method. Identifying the matched points with the SIFT algorithm is a procedure which is practically instant in comparison to the other parts of the program. The number of matched points that are identified, however, influences the number of systems of equations that must be solved, and this is a procedure which can be time consuming. In the case of the two images used in Fig. \ref{fig:fig2}, the entire program takes less than two minutes to identify points, solve for the proper angles, and plot the results on a standard personal computer (AMD Ryzen 5 3600X 6-Core processor); we believe that this time can be significantly reduced with more efficient algorithms. The program is believed to be very robust, as it was used for all 70 of the data points in Fig. \ref{fig:fig4} and also tested on an entirely different specimen. Only in 2 cases were an insufficient number of vectors formed, and this can likely be remedied by adjusting the constraints; in every other case a relatively large number of matched points were identified, and the resulting proper angles had a narrow distribution. If a sufficient number of points cannot be identified by the program, then hand measurements may be used if visible features exist that can be matched in both images; however, this may not be as accurate since SIFT identifies matched points down to a single pixel. The image pairs used for each datapoint in Fig. \ref{fig:fig4} were also of variable quality and brightness, so the program is generally robust even under a variety of conditions. The fact that the program requires only a small amount of input from the user (two SEM images and the tilt difference, which can be kept constant) means that it can be quickly and easily applied to any specimen; the increased accuracy and increased convenience makes this a valuable tool and one which should become standardized for this purpose.
 
Software packages other than the SIFT algorithm exist which identify correlated features between image pairs for different purposes. For example, 2D Digital Image Correlation (2D-DIC) uses a similar procedure and has been applied to deformation measurements in a variety of materials \cite{Sutton2007a, Sutton2007b}. The SIFT algorithm may also be suitable for this purpose, as it was a very effective tool in the present paper. For the purposes of both deformation measurement and specimen orientation, it is worthwhile to test different software packages, identify their strengths and weaknesses, and choose the one most suitable for the application at hand.

\section*{Conclusion}

We presented a first principles derivation of an in-situ method for use in a SEM that orients flat surfaces such that the surface normal is aligned with the electron beam. A program using the SIFT computer vision algorithm was developed to automate and expedite the process of making measurements on SEM images. The program was used to perform a detailed statistical analysis of the method on an experimental dataset and was able to detect errors in the SEM stage. Our results indicate that the specimen orientation method and the program to implement it have a high level of precision and can be used to quickly orient specimens so that quantitative analysis methods may be used.

\section*{Acknowledgements}

Financial support by National Science Foundation (DMR-1900077) is acknowledged. The authors also acknowledge Faith Corman, Joshua Homan, Brady Jones, and Abbeigh Schroeder for assistance in acquiring the SEM dataset.

\medskip

\noindent Competing interests: The author(s) declare none

\bibliographystyle{unsrt}
\bibliography{refs.bib}

\end{document}